# Bringing Information Credibility Back Into Transparency:
## The Case for a Global Monitoring System
## Of Green House Gas Emissions

– DRAFT Version –

Sébastien Philippe

sp6@princeton.edu
Program on Science and Global Security
Princeton University
May 10, 2016

**Abstract**
The goal of climate change governance is to stabilize greenhouse gas concentrations. This requires the reduction of anthropogenic global net emissions. In the pursuit of such a reduction, knowledge of greenhouse gas sources and sinks is critical to define baselines, and assess the effectiveness of climate governance over time. Such information and the means to independently verify its credibility continue to remain out of reach including in the recent Paris agreement. This essay argues that to make real progress in mitigating future climate change, this status quo must be challenged both intellectually and practically. First, it proposes to acknowledge and address the inconsistency between the objectives of a climate regime and the role of transparency as a mean to achieve these objectives. It does so by redefining transparency as the addition of publicity and measurability, which turns it into a credible information generating mechanism for governance. Second, it shows how in practice, a global monitoring system of greenhouse gas net emissions, based on this new definition of transparency and designed as a global public good, could provide the necessary knowledge to help frame governance solutions by providing both credibility and completeness in understanding the scope of the problem to solve.

## Introduction

In the ongoing climate negotiations within the United Nations Framework Convention on Climate Change (UNFCCC) that seek to stabilize atmospheric greenhouse gas (GHG) concentrations,[1] the Conference of the Parties has recently adopted the Paris Agreement and its objectives of "holding the global average temperature to well below 2 °C above pre-industrial levels and pursuing efforts to limit the temperature increase to 1.5 °C."[2]

---

[1] Article 2, United Nations Framework Convention on Climate Change (1994).
[2] Article 2, paragraph b of the Paris Agreement (2015).

To promote "effective implementation", the Agreement establishes "an enhanced transparency framework" to provide a clear understanding of climate change action.[3] Parties agreed to regularly provide national inventory reports of anthropogenic GHG emissions by sources and removals by sinks, as well as any information necessary to track their progress in implementing and achieving their nationally determined contribution to strengthen the global response to climate change.[4] This framework will be implemented "in a facilitative, non-intrusive, and non-punitive manner, respectful of national sovereignty, and avoid placing undue burden on Parties, " characteristic of a low level of international legalization.[5]

The historical record, so far, suggests that voluntary initiatives and self-reporting mechanisms are insufficient to induce compliance and effective participation.[6] This is particularly well established for regimes that need to create cooperation and deter free riding in order to achieve effective governance.[7] Furthermore, the information review and compliance mechanism of the Agreement based on experts review panels[8] seems rather weak if such panels have no access to independent means of verifying the information provided by the parties.[9]

This raises the question of how the transparency measures of the Agreement can be expected to be sufficient to achieve the objectives of the Agreement, and commit major emitting countries to GHG emissions limitations, in the first place when it seems to be theoretically unlikely.

The current scholarly debate on the implementation of climate measures that opposes top-down and bottom-up approaches seems to be partially responsible for this problem.[10,11] While both sides agree on the role of transparency as *publicity* that is an information creating and sharing mechanism, they disagree on how information credibility is assessed. Proponents of the top-down approach, which was historically pursued before Paris, often support a strong centralized monitoring, reporting and verification system. Advocates of a bottom-up strategy that has prevailed in the Paris Agreement, on the other hand, promote the emergence of clubs and experts' panels for information review but recognize at the same time that a useful unified mechanism has yet to emerge. Both sides, in their arguments, have come to separate the notion of information credibility from the notion of transparency.

---

[3] Article 13, paragraph b of the Paris Agreement (2015).
[4] Articles 2, 4 and 13 of the Paris Agreement (2015).
[5] Abbott, Kenneth W., Robert O. Keohane, Andrew Moravcsik, Anne-Marie Slaughter, and Duncan Snidal. "The concept of legalization." International organization 54, no. 03 (2000): 401-419.
[6] Peterson, M.J., "International Organizations and the implementation of Environmental Regimes," in Young, Oran R. (ed.) Global governance: drawing insights from the environmental experience. (MIT Press, Cambridge, MA, 1997). pp. 115-151.
[7] Barrett, Scott. Environment and statecraft: The strategy of environmental treaty-making. OUP Oxford, 2003.
[8] Article 13 paragraph 11 (transparency framework) provides that information submitted by each party shall undergo technical expert review. Article 15 establishes an expert based committee to facilitate implementation and promote compliance with the provisions of this Agreement.
[9] The most likely conclusion of such review panels seems rather trivial: data submitted are consistent with themselves.
[10] Hare, William, Claire Stockwell, Christian Flachsland, and Sebastian Oberthür. "The architecture of the global climate regime: a top-down perspective." Climate Policy 10, no. 6 (2010): 600-614.
[11] Sabel, Charles F., and David G. Victor. "Making the Paris process more effective: a new approach to policy coordination on global climate change." Policy Analsys Brief, The Stanley Foundation (2016): 1-8.

In this essay, I argue that to reach the objectives of the Paris Agreement as stated in Article 2, and of the regime complex for Climate Change more broadly,[12] climate scholars must re-integrate the notion of information credibility in transparency. This is the most compelling way to restore intellectual consistency while making practical efficacy possible. To do so, I propose to add the notion of *measurability* to the notion of *publicity* to redefine transparency.

This way, I argue that for a regime to be truly transparent,[13] the information it produces must be both public and measurable. Recognizing *measurability* as part of transparency provides the tools, to not only assess the completeness of the problem to solve, but also the information credibility necessary to make the debate and potential negotiations possible. I argue further that in the case of regimes that seek to govern physical global public goods, such as the atmosphere, measurability needs to be made itself a global public good to facilitate its collective governance.

Self-reporting alone satisfies the principle of publicity but not necessarily of measurability. In the case of global GHG emissions and sinks, independent data to support the evaluation and verification of states declarations or actions, i.e. the ability to attribute emissions and sinks to a state that can ensure a satisfactory level of treaty compliance, is still unavailable.[14] Yet, the means to acquire such data are currently an active area of research and development in the most technologically advanced countries.[15]

To satisfy the public good principle of measurability, governance and coordination of such means must be made global. The development of a global GHG monitoring system (GHG-GMS) from which measurability of GHG emissions and sinks would be globally available could satisfy such a requirement. I conclude that self-reported net emissions (publicity) together with a GHG-GMS (measurability) could provide the adequate level of transparency necessary for collective action. A GHG-GMS could emerge outside of the UNFCCC framework and be built from a variety of nodes following a building blocks strategy as advocated by Stewart, Oppenheimer and Rudyk.[16]

---

[12] Keohane, Robert O., and David G. Victor. "The regime complex for climate change." Perspectives on politics 9, no. 01 (2011): 7-23.

[13] Brunnée, Jutta, and Ellen Hey. "Transparency and International Environmental Institutions," in Peters, Anne, and A. Bianchi (eds). "Transparency in International Law." (Cambridge University Press, 2013): 23-48.

[14] Pacala, Stephen W. et al. "Verifying greenhouse gas emissions: Methods to support international climate agreements." Committee on Methods for Estimating Greenhouse Gas Emissions (2010).

[15] Ciais, Philippe et al. "Towards a European Operational Observing System to Monitor Fossil CO2 emissions." (Copernicus, European Commission, 1995): 1-68. Available at: http://www.copernicus.eu/sites/default/files/library/CO2_Report_22Oct2015.pdf

## Redefining transparency as publicity and measurability

It is known that transparency can play a fundamental role in treaty making and enforcement.[17,18,19] It is a notion difficult to define, however.[20] International environmental agreements (IAEs) that seek to regulate global public goods typically require the production of information in the form of reports, disclosure of activities or submission to observation and inspection.[21] Reporting generates information, while observation and inspection provide an assessment of its credibility. Chayes and Chayes suggest that the availability of such information measures the transparency of the regime.

In practice, transparency helps signal a party's commitment to an agreement before and after successful negotiations.[22] Its repetitive or continuous provision reassures other parties, and provides a basis to measure any departure from expected behavior making it a central mechanism to assess compliance and facilitate enforcement.[23] It also provides greater accountability of actors and institutions,[24] and supports the *Publicity Principle*.[25]

These descriptions hint toward a definition of transparency as the practice of making credible and valued information available to *enlighten* the state decision-making process. This assumes that critical information or knowledge for decision-making exists in the first place as well as the mean to assess its credibility.

For treaties that seek to regulate, limit, or ban a physical good, creating knowledge of that good requires its *measurability* (i.e. the capability to measure such good).

---

[16] Stewart, Richard B., Michael Oppenheimer, and Bryce Rudyk. "Building blocks for global climate protection." Stanford Environmental Law Journal 32, no. 2 (2013): 12-43.

[17] Mitchell, Ronald B. "Sources of transparency: Information systems in international regimes." International Studies Quarterly 42, no. 1 (1998): 109-130.

[18] Barrett, Scott. Environment and statecraft.

[19] Aldy, Joseph E. "The crucial role of policy surveillance in international climate policy." Climatic Change 126, no. 3-4 (2014): 279-292.

[20] Ball, Carolyn. "What is transparency?." Public Integrity 11, no. 4 (2009): 293-308.

[21] Chayes, Abram, and Antonia Handler Chayes. "Compliance without enforcement: state behavior under regulatory treaties." Negotiation Journal 7, no. 3 (1991): 311-330.

[22] Schelling, Thomas C. "An essay on bargaining." The American Economic Review (1956): 281-306.

[23] Simmons, Beth A. "Compliance with international agreements." Annual Review of Political Science 1, no. 1 (1998): 75-93.

[24] Ebbesson, Jonas. "Global or European only? International law on transparency in environmental matters for members of the public," in Peters, Anne, and A. Bianchi (eds). "Transparency in International Law." (Cambridge University Press, 2013): 49-74.

[25] Luban, David. "The publicity principle." In Robert E. Goodin, The theory of institutional design (Cambridge University Press, 1996): 154-198.

The credibility of information is obtained from the credibility of the measurement or observation that can be established, for example, by its independent reproducibility. In the context of regime governance, measurability is necessary for verifiability.[26]

Consequently, the timeline between technology, knowledge and action is crucial in treaty negotiation. When critical information is lacking, essential players may be unwilling or unable to coordinate their behavior.[27] If knowledge is required for action, and technology is required for knowledge, then technology must be available to trigger action. The absence of technology to establish knowledge only supports two outcomes: inaction, for example the inability of countries to cooperate and abate collectively GHG emissions since 1992, or inadequate governance, for example attempts at fisheries regulation when species critical population level (the minimum number of individuals to sustain a population) are unknown.[28]

It is therefore fundamental to seek the development of knowledge-creating technology when it is unavailable. Technology is not a static object, however, it continuously evolves and ongoing negotiations or established regimes are affected by new developments. If scientific knowledge can be considered a global public good,[29] a single actor can have an important impact on technological development. However, technology itself is not a public good and, for well-known reasons, states may have little incentives to share their technological advancements.

What are the implications for regime transparency? First, transparency requires measurability because it provides evidence that the information made public is credible. In the context of a court, this would correspond to proving the facts alleged. Any fact not proven will not to be taken into account by a tribunal: *idem est non probari non esse* (Riddell and Plant 2009).[30] Second, measurability must be made available to all parties, individually, or – at the bare minimum – to the parties that do not accept public information, in particular self-reported information, *prima facie* from one another but are trusted by others.[31] Third, the existence of a trusted institution with global membership as part of the regime can provide a substitute to individual needs of measurability.

---

[26] Breidenich, Clare, and Daniel Bodansky. Measurement, reporting and verification in a post-2012 climate agreement. Washington, DC: Pew Center on Global Climate Change, 2009.
[27] Sabel, Charles F., and David G. Victor. "Making the Paris process more effective: a new approach to policy coordination on global climate change."
[28] Barrett, Scott, and Astrid Dannenberg. "Climate negotiations under scientific uncertainty." Proceedings of the National Academy of Sciences 109, no. 43 (2012): 17372-17376.
[29] Dalrymple, Dana. "Scientific knowledge as a global public good: Contributions to innovation and the economy." In The Role of Scientific Data and Information in the Public Domain: Proceedings of a Symposium, pp. 35-51. National Academies Press, 2003.
[30] Something that is not proven does not exist.
[31] Here, I take a neoliberal institutionalist view of *trust* in international relation. According to Hoffman (2002): "*trust* in international relation refers to an attitude involving a willingness to place the fate of one's interests under the control of others. This willingness is based on a belief, for which there is some uncertainty, that potential trustees will avoid using their discretion to harm the interests of the first." Hoffman, Aaron M. "A conceptualization of trust in international relations." European Journal of International Relations 8, no. 3 (2002): 375-401.

## Turning Measurability into a Global Public Good

If publicity and measurability are a response to the absence of trust, we can postulate that they must be made available as widely as possible to facilitate the construction of multilateral regimes. In the case where the regime purpose is to govern a physical global public good, for example the atmosphere, then the means to measure that public good must be made itself a global public good to facilitate its governance.[32] These means must themselves be trusted – if they are to be used to perform independent data analysis.

We can classify independent means of measurement in three forms: First, they can be single state-owned goods. In this case, they are often refereed as *national technical mean*s (NTMs) – a term coined during arms control negotiations between the US and USSR.[33] An agreement where measurability is provided by NTMs assumes a certain level of parity and reciprocity in the means of measurement. As the number of parties to a treaty increases, the probability of establishing parity – that is the availability of measurability to all party individually – certainly decreases because, as mentioned, means of measurement may require advanced technology such as space-based sensors and large resources. This is particularly problematic when an international environmental agreement seeks to govern a global public good. Second, means of measurement can be club goods or regional public goods. For example, member states of the European union can access shared spaced-based instruments through the European space agency. However, in this case, means of measurement are excludable and subject to artificial scarcity. This can create an incentive for membership in the club, however, and reduce accessibility cost to measurement resources. Finally, means of measurement can be global public goods. Here, both the means of measurement and data produced are fully and openly available through a global institution or an overlapping network of clubs that would achieve global (or near global) membership in aggregate[34]. The World Meteorological Organization,[35] and the international monitoring system of the Comprehensive Nuclear-Test-Ban Treaty (CTBT)[36] are two important examples. Only in this last case, parity and reciprocity are achieved for all, providing a fair and non-discriminatory access to independent verification, two important principles in international law making.[37]

---

[32] A global public good must be non-rival in consumption, non-excludable and a global good.
[33] Krass, Allan. "Verification: How Much Is Enough." Stockholm: SIPRI/Lexington Books (1985): pp 271.
[34] Casella, Alessandra, and Bruno Frey. "Federalism and clubs: Towards an economic theory of overlapping political jurisdictions." European Economic Review 36, no. 2 (1992): 639-646.
[35] Soroos, Marvin S., and Elena N. Nikitina. "The World Meteorological Organization as a purveyor of global public goods." Contributions In Political Science 355 (1995): 69-82.
[36] United Nations General Assembly, "Comprehensive Nuclear-Test-Ban Treaty," (1992), text available at: https://www.ctbto.org/fileadmin/content/treaty/treatytext.tt.html

## Creating a Global Monitoring Network of Green House Gas Emissions

In the pursuit of collective GHG net emission reduction, knowledge of GHG sources and sinks is critical to define baselines, and assess the effectiveness of climate governance over time. Current international negotiations have so far relied on self-reported data to the UNFCCC. Such data are most of the time based on proxies. For example, $CO_2$ emissions are usually based on energy-use statistics collected from various sources in different sectors of the economy.[38] There are no truly independent data against which data in self-declared inventory can be compared.

Several international or national institution such as the United Nations, the International Energy Agency, or the United States' Department of Energy (DOE) Energy Information Administration, create large international datasets on energy production and consumption, but all of these datasets rely primarily on the same self-reported national statistics. In many countries, these data are not complete or accurate or are not consistently reported.[39,40] Recent controversies about whether or not China's emissions have potentially peaked,[41] and if the country is burning more coal than it is reporting,[42] highlights the problem of continuously relying on self-reported data-proxies.

This problem is likely to continue for the foreseeable future, however. The UNFCCC principle of "common but differentiated responsibilities" – also an oxymoron – has lead states to declare in 2015 their national emission pledges as Intended Nationally Determined Contributions (INDCs), which have become the basis of the Paris Agreement and its "enhanced framework" for transparency. Obviously this framework doesn't have the means of measurability required to assess the credibility of public information declared by individual countries.

---

[37] Franck, Thomas M., and Thomas M. Franck. Fairness in international law and institutions. Oxford: Clarendon Press (1995).
[38] Ciais, Philippe et al. "Towards a European Operational Observing System to Monitor Fossil CO2 emissions."
[39] US Government Accountability Office, Literature on the Effectiveness of International Environmental Agreements, RCED-99-148 (1999).
[40] Pacala, Stephen W. et al. "Verifying greenhouse gas emissions: Methods to support international climate agreements."
[41] Wong, Edward. "China's Carbon Emissions May Have Peaked, but It's Hazy." *The New York Times*, April 3, 2016, available at: http://www.nytimes.com/2016/04/04/world/asia/china-climate-change-peak-carbon-emissions.html?_r=0
[42] Buckley, Chris. "China Burns Much More Coal Than Reported, Complicating Climate Talks," *The New York Times,* November 3, 2015, available at:
http://www.nytimes.com/2015/11/04/world/asia/china-burns-much-more-coal-than-reported-complicating-climate-talks.html

A solution to address this important problem could be to develop a global monitoring system of GHG (GHG-GMS) net emissions. Such a system would be based on the transparency principles of publicity and measurability and should be designed as a global public good.

A GHG-GMS has the potential to enable the accurate, transparent and consistent quantification of GHG net emissions and their trends at multiple spatial scales.[43] It can provide data for effective local and regional climate and energy policies, independent means of measurement of GHG net emissions for countries as part as climate agreements, and finally, assess aggregate collective efforts to mitigate climate change as well as provide important data for the governance of the atmosphere – a global public good.

Atmospheric measurements provided by a GHG-GMS could be used in an operational system that could comprise three complementary components:
- Atmospheric measurements obtained from dedicated space-based sensors complemented by ground-based (in-situ) networks,
- The provision of bottom-up information (for example fossil fuel consumption) available by linkage through the UNFCCC reporting process and other relevant institutions (potentially with higher spatial and temporal resolution),
- A trusted and certified data-processing center that will integrate both types of information into consistent, accurate, and independently verified estimates of GHG net emissions.

A GHG-GMS could emerge from the integration of existing scientific national means and transnational collaborations following a building block or bottom-up strategy.[44] Space-based systems have experienced important development recently. Existing, planned and potential missions (focusing mainly on $CO_2$) include OCO-2, OCO-3, and potentially GEOstationary (all operated by NASA) in the United States, MICROCARB in France, TanSat in China, GOSAT-1 and -2 in Japan, and Proposed Copernicus missions in the EU.[45] All these countries are part of the largest emitters. International data exchange agreements from these different missions would facilitate the development of a global integrated system. In addition, a GHG-GMS could benefit from emerging private actors in the space sector that could have interest in developing large constellations of sensors (for example SpaceX and Google Terra Bella among others).

---

[43] This section draws in part from the proposal of a European Union integrated monitoring system in: Ciais, Philippe et al. "Towards a European Operational Observing System to Monitor Fossil CO2 emissions."

[44] Stewart, Richard B., Michael Oppenheimer, and Bryce Rudyk. "Building blocks for global climate protection."

[45] Ciais, Philippe et al. "Towards a European Operational Observing System to Monitor Fossil CO2 emissions."

On the ground, important networks for atmosphere monitoring include ICOS (Integrated Carbon Observation System, Europe-based), NOAA (global, US-based), TCCON (Total Carbon Column Observing Network, multinational), AGAGE (Advanced Global Atmospheric Gases Experiment, US-funded), FLUXNET (network of regional networks, US-based), and GAW (Global Atmospheric Watch, operated by the World Meteorological Organization).[46]

The integration and further development of both space and ground components could be promoted by the World Meteorological Organization (WMO), the specialized United Nations agency for meteorology (weather and climate) that benefits from a large membership. The WMO has played a major role in climate negotiations by creating, together with the United Nations Environment Programme (UNEP), the Intergovernmental Panel on Climate Change (IPCC) that is the scientific advisory body of the UNFCC process.

The WMO has been recently developing plans for an Integrated Global Greenhouse Gas Information System (IG3IS). Such a system seeks to combine "ground-based and space-based observations, carbon-cycle modeling, fossil fuel-use and land-use data, meta-analysis, and an extensive distribution system to provide information about sources and sinks of greenhouse gases at policy-relevant temporal and spatial scales." This system could be the seed to develop a GHG-GMS as a global public good. The IG3IS is "envisioned as an independent, observationally based information system for determining trends and distributions of GHGs in the atmosphere and the ways in which they are consistent or not with efforts to reduce greenhouse gas emissions." It is not meant to statute on international agreements compliance but can provide the necessary transparency to do so.[47]

During, its last 2015 meeting, the WMO congress has requested its members:[48] (1) To give all possible support to the development, improvement and modernization of networks for observations of greenhouse gases and co-emitted species; (2) To ensure submission of observational data as well as metadata to the dedicated WMO/GAW Data Centre as well as the GAW Station Information System to support IG3IS; (3) To collaborate with organizations and institutions that address the carbon budget of biosphere and ocean.

The WMO has an important advantage over the UNFCCC for developing a GHG-GMS. Decisions are voted by two-thirds majority of the vote of a congress formed by its member states,[49] while the UNFCCC decisions require consensus.

---

[46] Tarasova, Oksana and James Butler. "Implementing an Integrated, Global Greenhouse Gas Information System (IG3IS)." Towards a Global Carbon Observing System: Progress and Challenges, 1-2 October 2013, Geneva.

[47] The Integrated Global Greenhouse Gas Information System, World Meteorological Organization, information retrieved at: https://www.wmo.int/pages/prog/arep/gaw/ghg/IG3IS-info.html

[48] Seventeenth World Meteorological Congress, Geneva, 25 May–12 June 2015, Abridged final report with resolutions. Accessed at: http://www.cmoc-china.cn:8080/upfiles/pdf/wmo_1157_en.pdf

[49] Article 10b, Convention of the World Meteorological Organization (1947).

The combination of emergent sensors networks operated by individual nations or clubs (potentially involving private actors) and the potential of the WMO to host and coordinate the growth of a GHG-GMS is surprisingly promising and shall be encouraged in a more vocal way, perhaps by getting support and traction from civil society. This approach has the benefit of providing measurability for the climate regime, while guarantying a fair access to valuable information for developing and smaller states. It could continue to evolve in the future to provide support for the potential regulation of atmospheric geoengineering.

This solution does require the technological support and financing of a handful of nations that are also the largest emitters and may be reluctant to provide detailed verified information about their emissions.

There is hope, however, as all majors GHG emitters are members of treaties that involve measurability sometimes in a very intrusive manner. Six relevant examples are briefly presented in the appendix. They show precedent of large area monitoring involving aerial photography, international network of sensors, limitations and measurements of atmospheric concentrations of regulated substances as well as challenge inspections in these countries.

## Conclusion

Knowledge of greenhouse gas sources and sinks is critical to define baselines, and assess the effectiveness of climate governance over time. Such information and the means to independently verify its credibility continue to remain out of reach including in the recent Paris agreement. To make real progress in mitigating future climate change, I argue that this status quo must be challenged both intellectually and practically.

A first necessary step in that direction is to acknowledge and address the inconsistency between the objectives of a climate regime and the role of transparency as a mean to achieve these objectives. In this essay, I did so by redefining transparency as the addition of publicity and measurability, which turns it into a credible information generating mechanism.

Second, I have shown how a global monitoring system of greenhouse gas net emissions based on this definition of transparency as publicity and measurability and designed as a global public good could provide the necessary knowledge to help frame governance solutions by providing both credibility and completeness in understanding the scope of the problem to solve.

In practice, the prospect of combining emergent sensors networks operated by individual nations or clubs, and the potential of the World Meteorological Organization to coordinate and govern a global monitoring system of the atmosphere, including greenhouse gas net emissions, is surprisingly promising and would benefit from further research.

# Appendix. Accepting measurability: some evidence from existing multilateral treaties involving large GHG emitters

This appendix briefly presents six cases in which major GHG emitters have accepted measurability in multilateral treaties, sometimes in a very intrusive manner. Here, I consider the following countries or entities: China, the United States, the European Union (28), India, the Russian Federation, Indonesia, Brazil, Japan, Canada, and Mexico. These treaties are concerned with the limitation or ban of physical substances or activities – or of global information gathering such as general military activities. They all allow for measurements, observations and potentially inspections, all relatively intrusive mechanism of independent information gathering.

These treaties are:

1. **The Treaty on Open Skyes** (US, EU, Russian Federation and Canada). The treaty establishes a program of unarmed aerial surveillance flights over the entire territory of its participants.
2. **Comprehensive Nuclear Test Ban Treaty** (EU, Indonesia, Brazil, Japan, Canada, Mexico have signed and ratified. US, Russian Federation and China have signed). The treaty establishes an international monitoring system for the detection of nuclear explosion. It consists of a global network of sensors including seismographs, hydrophones, infrasound microphones, and radionuclides air sampling stations.
3. **MARPOL 73/78**, The International Convention for the Prevention of Pollution from Ships (All are members): provides the right to inspect ships at port and detain them in cases of non-compliance. It introduced air pollution standards in 2005.
4. **Chemical Weapons Convention** (All have signed and ratified): Article V, Par. 7b provides the right to conduct challenge inspection anywhere in the entire territory of participants.
5. **Montreal Protocol** (All have signed and ratified): monitors regulated substances atmospheric concentration although not for individual countries.
6. **Convention on Long-Range Transboundary Air Pollution** (EU, US, Russian Federation and Canada): sets up the Co-operative Programme for Monitoring and Evaluation of the Long-range Transmission of Air Pollutants in Europe that relies on collection of emission data, measurements of air and precipitation quality.